\documentclass[preprint,aps]{revtex4}

\usepackage{graphicx}
\usepackage{dcolumn}
\usepackage{bm}

\begin{document}

\title{Bi-Large Neutrino Mixing See-Saw Mass Matrix with Texture Zeros and Leptogenesis}

\author{$^1$Wei Chao\footnote{chaowei@mail.nankai.edu.cn}, $^{1,2}$Xiao-Gang He\footnote{hexg@phys.ntu.edu.tw},
 and $^1$Xue-Qian Li\footnote{lixq@nankai.edu.cn}}
\affiliation{%
$^1$ Department of Physics, Nankai University, Tianjin\\
$^2$NCTS/TPE, Department of Physics, National Taiwan University,
Taipei
}

\date{\today} 

\begin{abstract}
We study constraints on neutrino properties for a class of
bi-large mixing See-Saw mass matrices with texture zeros and with
the related Dirac neutrino mass matrix to be proportional to a
diagonal matrix of the form $diag(\epsilon, 1,1)$. Texture zeros
may occur in the light (class a)) or in the heavy (class b))
neutrino mass matrices. Each of these two classes has 5 different
forms which can produce non-trivial three generation mixing with
at least one texture zero. We find that two types of texture zero
mass matrices in both class a) and class b) can be consistent with
present data on neutrino masses and mixing. None of the neutrinos
can have zero masses and the lightest of the light neutrinos has a
mass larger than about 0.039 eV for class a) and $0.002$ eV for
class b). In these models although the CKM CP violating phase
vanishes, the non-zero Majorana phases can exist and play an
important role in producing the observed baryon asymmetry in our
universe through leptogenesis mechanism. The requirement of
producing the observed baryon asymmetry can further distinguish
different models and also restrict the See-Saw scale to be in the
range of $10^{12}\sim 10^{15}$ GeV.

\end{abstract}

\maketitle

\newpage

\section{Introduction}

There are abundant experimental data showing that neutrinos have
small but non-zero masses and also mix with each
other\cite{neutrino1,neutrino2,neutrino3,neutrino4,neutrino5,data,pdg}.
One of the most interesting mechanisms of naturally generating
small neutrino masses is the See-Saw mechanism\cite{see-saw}. This
mechanism requires introduction of right-handed neutrinos $\nu_R$
into the theory. At a phenomenological level, the See-Saw neutrino
mass matrix, in the left-handed neutrino $\nu_L$ and the charge
conjugated right-handed neutrino $\nu^c_R$ basis
($\nu_L,\;\;\nu^c_R$) with the charged lepton mass matrix already
diagonalized, can be written as
\begin{eqnarray}
M = \left ( \begin{array}{cc}
0&M_D^T\\
M_D&M_R\end{array}\right ),
\end{eqnarray}
where $M_D=v_\nu Y_\nu$ is the Dirac neutrino mass term which can
be generated through the Yukawa couplings of a Higgs doublet
$H_\nu$ to the left- and right-handed neutrinos, $\bar \nu_R
(Y_\nu v_\nu) \nu_L$ with $v_\nu$ being the vacuum expectation
value of $H_\nu$. $M_R$ is from the Majorana mass term $(1/2)\bar
\nu_R M_R \nu^c_R$.

With three generations of left- and right-handed neutrinos, $M_R$
is a $3\times 3$ symmetric matrices, and $M_D$ is a $3\times 3$
arbitrary matrix. The elements in $M_D$ can be of the same order
of the magnitude as the corresponding charge leptons, and the
scale of $M_R$ is a new scale characterizing possible new physics
beyond SM which is expected to be much larger than the weak scale.
To the leading order, the mass matrices $M_h$ and $M_\nu$ for the
heavy and light neutrinos are given by

\begin{eqnarray}
M_h \approx M_R,\;\; M_\nu \approx - M_D^T M^{-1}_R M_D =
-v_\nu^2 Y_\nu^T M^{-1}_{R} Y_\nu.
\end{eqnarray}

The light neutrino masses are suppressed compared with their
charged lepton partners by a factor of $M_D/M_R$ resulting in very
small neutrino masses compared with the masses of their
corresponding charged leptons. The eigen-values and eigen-vectors
of $M_\nu$ are the light neutrino masses and mixing measured by
low energy experiments. The mixing matrix is the unitary matrix
$U$ (the PMNS matrix\cite{mns}) which diagonalizes the mass
matrix, and is defined, in the basis where the charged lepton is
already diagonalized, by $D = U^T M_\nu U = diag(m_{1}, m_{2},
m_{3})$ with the eigenvalues $m_i$ to be larger or equal to zero.
One can always decompose $U$ into a product of a CKM
matrix\cite{ckm} like unitary matrix $V$ and a phase matrix $P =
diag(e^{i\rho_1}, e^{i\rho_2}, e^{i\rho_3})$, $U = VP$. The phase
$\rho_i$ is the Majorana phase. It is some times convenient to
write $\tilde D = V^T M_\nu V$. In this case the eigenvalues are
in general complex which will be indicated by $\tilde m_i = m_i
e^{-i2\rho_i}$.  The commonly used parametrization for $V$ is
given by\cite{pdg}
\begin{eqnarray}
V = \left (\begin{array}{ccc}
1&0&0\\
0&c_{23}&s_{23}\\
0&-s_{23}&c_{23}
\end{array}
\right )
 \left (\begin{array}{ccc}
c_{13}&0&s_{13}e^{i\delta}\\
0&1&0\\
-s_{13}e^{-i\delta}&0&c_{13}
\end{array}
\right )
 \left (\begin{array}{ccc}
c_{21}&s_{21}&0\\
-s_{21}&c_{12}&0\\
0&0&1
\end{array}
\right ),
\end{eqnarray}
where $s_{ij} = \sin\theta_{ij}$ and $c_{ij} = \cos\theta_{ij}$.
$\delta$ is a CKM like CP violating phase.

If See-Saw mechanism is responsible for neutrino masses and
mixing, the properties of the right-handed neutrinos play a very
important role in determining the light neutrino properties. $M_R$
not only provides a scale for new physics responsible for the
mechanism to explain the smallness of neutrino masses, but also
affects the low energy mixing, and vice versa. It may also provide
important ingredients to explain the baryon asymmetry of our
universe (BAU) through lepton number violating decays of the heavy
neutrino to light neutrinos and Higgs particles by the Yukawa
coupling $Y_\nu$, the leptogenesis
mechanism\cite{fukugita,leptogenesis,asy}. If one takes
leptogenesis as a requirement, important information about the
mass matrix and the associated CP violating phases can be
obtained\cite{grimus2,leptocp}. The CP violating Majorana phases
$\rho_i$ can play an important role in explaining BAU through
leptogenesis mechanism which will be discussed later.

The present neutrino oscillation experimental data on neutrino
masses and mixing angles
from\cite{neutrino1,neutrino2,neutrino3,neutrino4,neutrino5}
solar, atmospheric, reactor neutrino oscillation experiments can
be summarized as the following.  The $3\sigma$ allowed ranges for
the mass-squared differences are constrained to be\cite{data,pdg}:
$1.6 \times 10^{-3}$ eV$^2$ $\leq \Delta m^2_{atm} = |\Delta
m^2_{32}| = |m^2_3-m^2_2|\leq 3.6\times 10^{-3}$ eV$^2$, and
$7.3\times 10^{-5}$ eV$^2$ $\leq \Delta m^2_{solar} = \Delta
m^2_{21}= m^2_2 - m^2_1 \leq 9.3 \times 10^{-5}$ eV$^2$, with the
best fit values given by $\Delta m^2_{atm} = 2.2\times 10^{-3}$
eV$^2$, and $\Delta m^2_{solar} = 8.2\times 10^{-5}$ eV$^2$. The
mixing angles are in the ranges of $0.28\leq \tan^2\theta_{12}
\leq 0.60$ (best fit value 0.39), $0.5 \leq \tan^2\theta_{23} \leq
2.1$ (best fit value 1.0), and $\sin^2\theta_{13} \leq 0.041$.

There are many theoretical studies of neutrino masses and
mixing\cite{matrix}. The mixing matrix can be nicely represented
by the so called bi-large mixing matrix. By an appropriate choice
of sign and phase conventions, the bi-large mixing matrix can be
written as
\begin{eqnarray}
V = \left (\begin{array}{rrr}
c&s&0\\
-{s\over \sqrt{2}}&{c\over \sqrt{2}}&{1\over \sqrt{2}}\\
-{s\over \sqrt{2}}&{c\over \sqrt{2}}&-{1\over \sqrt{2}}
\end{array}\right ),
\end{eqnarray}
where $c = \cos \theta$ and $s=\sin\theta$ with $\tan\theta
\approx \tan\theta_{12}$. In this form the eigen-masses $\tilde
m_{\nu_i}$ of the light neutrinos, in general, have non-zero
Majorana phases $\rho_i$.

This class of models has $V_{13}=0$ which is allowed by present
experimental data and can be tested by future experiments. Several
experiments are planned to measure $V_{13}$ with greater
precisions\cite{daya}. Obviously should a non-zero value for
$V_{13}$ be measured, modifications for the model considered are
needed. The bi-large mixing model can be taken as the lowest order
approximation. The bi-large mixing model captures many features of
the present data and deserves more careful theoretical studies.

The bi-large mixing mass matrix is of the
form\cite{matrix,grimus1}

\begin{eqnarray}
M_\nu &=& \left(   \begin{array}{ccc}
       M_{11} & M_{12} & M_{12} \\
       M_{12} & M_{22} & M_{23} \\
       M_{12} & M_{23} & M_{33} \\
  \end{array}   \right)\nonumber\\
&=& \left ( \begin{array}{lll} (c^2\tilde m_{\nu_1} + s^2 \tilde
m_{\nu_2})& {cs\over \sqrt{2}}(-\tilde m_{\nu_1} +\tilde
m_{\nu_2})&
{cs\over \sqrt{2}}(- \tilde m_{\nu_1} + \tilde m_{\nu_2})\\
{cs\over \sqrt{2}}(-\tilde m_{\nu_1} + \tilde m_{\nu_2})&
{1\over 2}(s^2 \tilde m_{\nu_1} + c^2 \tilde m_{\nu_2} + \tilde m_{\nu_3})&
{1\over 2}({s^2 \tilde m_{\nu_1}} + {c^2 \tilde m_{\nu_2}} -{ \tilde m_{\nu_3}})\\
{cs\over \sqrt{2}}(-{ \tilde m_{\nu_1}} + { \tilde m_{\nu_2}})& {1\over 2}
({s^2\tilde m_{\nu_1}} + {c^2\tilde m_{\nu_2}} -{\tilde m_{\nu_3}})&
{1\over 2}({s^2 \tilde m_{\nu_1}} + {c^2 \tilde m_{\nu_2}} +{\tilde m_{\nu_3}})
\end{array} \right ).
\label{mnu}
\end{eqnarray}

The above mass matrix produces bi-large mixing matrix, but the
neutrino masses cannot be completely fixed using the two mass
different $\Delta m^2_{sol}$ and $\Delta m^2_{atm}$ measurements
even one knows all information about the mixing matrix. Additional
inputs are needed to further constrain or to determine the
parameters in the neutrino mass
matrix\cite{matrix,grimus1,grimus,texture,determinant,traceless}.
Several proposals have been made to reduce the parameters, such as
texture zero\cite{texture}, determinant zero
requirement\cite{determinant}, and traceless
requirement\cite{traceless} for the mass matrix. In this paper we
impose texture zeros on bi-large mixing mass matrix to constrain
neutrino masses and Majorana phases and also study the
implications for leptogenesis.

To study leptogenesis, one needs further information on the heavy
neutrino mass matrix. Since $M_\nu =- M_D M^{-1}_R M_D^T$, if
$M_D$ is known one can obtain the form of $M_R$ at higher energy.
If $M_D$ is proportional to a unit matrix, $M_R$ will also have a
similar form as the one given in eq.(\ref{mnu}), and the
associated mixing matrix $V_R P_R$ is equal to $V^*P^*$ in the
basis where all eigen-masses are real and positive, and the heavy
neutrino eigen-masses are proportional to $1/m_i$. $M_\nu$ and
$M_R$ are trivially related. We will show later that in this case
the Majorana phases will play no role in leptogenesis. We
therefore consider a simple, but non-trivial relation between
$M_\nu$ and $M_R$ with $M_D = v_\nu Y_\nu = v_\nu b\;
diag(\epsilon ,1,1)$. With this form for $M_D$, $M_R$ also has the
bi-large mixing mass matrix form, but with $V_R P_R$ not equal to
$V^*P^*$. With these forms for $M_\nu$ and $M_R$, there is no CKM
like CP violating phase in the charged current interaction with
$W$ boson. The Majorana phases can, however, play a non-trivial
role in leptogenesis\cite{leptocp,grimus2,majoranaphase} and will
be studied in more details later. The heavy neutrino mass matrix
can be expressed as

\begin{eqnarray}
M_R = -v_\nu^2 b^2 \left ( \begin{array}{lll}
\epsilon^2 ({c^2\over \tilde m_{\nu_1}} +
{s^2\over \tilde m_{\nu_2}})& \epsilon {cs\over \sqrt{2}}(-{1\over \tilde m_{\nu_1}}
+{1\over \tilde m_{\nu_2}})&
\epsilon {cs\over \sqrt{2}}(-{1\over \tilde m_{\nu_1}} + {1\over \tilde m_{\nu_2}})\\
\epsilon {cs\over \sqrt{2}}(-{1\over \tilde m_{\nu_1}} + {1\over \tilde m_{\nu_2}})&
{1\over 2}({s^2\over \tilde m_{\nu_1}} + {c^2\over \tilde m_{\nu_2}} +{1\over \tilde m_{\nu_3}})&
{1\over 2}({s^2\over \tilde m_{\nu_1}} + {c^2\over \tilde m_{\nu_2}} -{1\over \tilde m_{\nu_3}})\\
\epsilon {cs\over \sqrt{2}}(-{1\over \tilde m_{\nu_1}} + {1\over \tilde m_{\nu_2}})& {1\over 2}
({s^2\over \tilde m_{\nu_1}} + {c^2\over \tilde m_{\nu_2}} -{1\over \tilde m_{\nu_3}})&
{1\over 2}({s^2\over \tilde m_{\nu_1}} + {c^2\over \tilde m_{\nu_2}} +{1\over \tilde m_{\nu_3}})
\end{array} \right ).
\label{mr}
\end{eqnarray}
We note that none of the neutrino masses can be zero.

We comment that when discussing See-Saw neutrino mass matrices,
there are two scales, the light neutrino and the heavy neutrino
mass scales. The mass matrices at the two scales may be different
due to renormalization group running effects\cite{running}. A mass
matrix element is zero at a particular scale may not be zero at
another scale unless there are certain symmetries to guarantee
this\cite{screening}.

 An important problem in neutrino physics is to understand
the origin of neutrino masses and mixing. There are many attempts
have been made to understand this problem, at present the answer
is, however, far from satisfaction. We will not attempt to carry
out a model building investigation of the mass matrices, instead
we will study phenomenological implications here. However, we
would like to point out there are models which can produce some
models we study. For example, in Ref\cite{grimus} it was shown
that for three generations of left-handed leptons and three
generations of right-handed charged leptons and neutrinos, it is
possible to obtain $M_D \sim diag(\epsilon, 1,1)$ and $M_R$ of the
form in eq.(\ref{mr}) with $M_{23} = 0$ if there are 3 Higgs
doublets and 2 neutral signets which transform, non-trivially,
under a discrete group $Z_2^{(\tau)}\times Z^{(tr)}_2\times
Z^{(aux)}_2$ . In this model the flavor symmetries of the model
dictates that the light neutrino mass matrix structure is
essentially determined by the heavy neutrino mass
matrix\cite{screening}.

We will generalize the discussion to include all possible texture
zeros in both the light and heavy bi-large mass matrices. Although
the classes of models have simple structures, there are very rich
phenomenological implications on the neutrino masses, CP violating
phases and on leptogenesis.

\section{Bi-Large Neutrino Mixing mass matrix with texture zeros}

There are two different ways the texture zeros can be imposed: a)
The texture zeros are imposed on the light neutrino mass matrix
$M_\nu$; And b) The texture zeros are imposed on the heavy
neutrino mass matrix $M_R$. There are five different cases for
each of class a) and class b) types of models which give
non-trivial three generation mixing.

For class a) the five cases indicated by $M_{Li}$ are
\begin{eqnarray}
&&M_{L1} = \left(    \begin{array}{ccc}
0 & M_{12} & M_{12} \\
M_{12} & M_{22} & M_{23} \\
M_{12} & M_{23} & M_{22} \\
\end{array}    \right),
\;M_{L2}= \left(    \begin{array}{ccc}
0 & M_{12} & M_{12} \\
M_{12} & 0 & M_{23} \\
M_{12} & M_{23} & 0 \\
\end{array}    \right),\;
M_{L3} = \left(    \begin{array}{ccc}
0 & M_{12} & M_{12} \\
M_{12} & M_{22} & 0 \\
M_{12} & 0 & M_{22} \\
\end{array}    \right),\nonumber\\
&&M_{L4} = \left(    \begin{array}{ccc}
M_{11} & M_{12} & M_{12} \\
M_{12} & M_{22} & 0 \\
M_{12} & 0 & M_{22} \\   \end{array}    \right),\;
M_{L5} = \left(   \begin{array}{ccc}      M_{11} & M_{12} & M_{12} \\
M_{12} & 0 & M_{23} \\       M_{12} & M_{23} & 0 \\   \end{array}    \right).
\end{eqnarray}

For class b) the five cases have the same forms as above. We
denote them and their matrix elements by $M_{Ri}$ and $M^h_{ij}$,
respectively.

Before discussing the above mass matrices in detail, we make some
comments on the neutrino masses. There are three light neutrino
masses, and there are two observable mass differences $\Delta
m^2_{sol}$ and $\Delta m^2_{atm}$ from neutrino oscillation. If
one of the neutrino masses, or a combination of them, is known,
the rest of the masses can be determined in terms of $\Delta
m^2_{21,32}$. We express $m_2$ and $m_3$ as a function of $m_1$
as,
\begin{eqnarray}
m_{\nu_2} = \sqrt{m^2_{\nu_1} +\Delta m^2_{21}}, \;\;m_{\nu_3} =
\sqrt{m^2_{\nu_1} + \Delta m_{21}^2+\Delta m^2_{32}}.
\end{eqnarray}

There are additional constraints on the neutrino masses. One of
them comes from WMAP data which limits $m_{sum} = m_1 + m_2 + m_3$
to be less than\cite{wmap} $ 0.71$ eV at the 95\% C.L.. In the
future cosmological data can improve the sensitivity and reach
0.03 eV\cite{weiler}. There are another two constraints, the
effective mass for neutrinoless double beta decays, $m_{ee} =
|M_{11}|$, and the effective mass for tritium beta decay, $\langle
m_{\nu}\rangle = (|V_{11}|^2 m^2_1 + |V_{12}|^2 m^2_2 + |V_{13}|^2
m^2_3)^{1/2}$. The current experimental upper bounds for $m_{ee}$
and $\langle m_\nu \rangle$ are 1.35 eV\cite{doublebeta} and 3
eV\cite{pdg}, respectively. These bounds are less stringent than
the WMAP constraint. However, in the future, the sensitivities for
$m_{ee}$ and $\langle m_\nu\rangle$ can reach 0.01 eV
\cite{doblebeta1} and 0.12 eV\cite{tritium}, respectvely, by
laboratory experiments. These experiments can provide interesting
constraints.

If there are additional constraints, such as texture zeros, it may
be possible to determine the neutrino masses or relate the masses
to the CP violating Majorana phases which are otherwise very
difficult to measure.

\subsection{Constraints From Texture Zeros In The Light Neutrino Mass
Matrix}

In this section we study the consequences of the above texture
zeros in class a). For the case L1, we have
\begin{eqnarray}
M_{11} = c^2 \tilde m_1 + s^2 \tilde m_2 = 0.
\end{eqnarray}
This is a special case studied in Ref.\cite{xing} by requiring
$m_{ee}=0$.

 The phase of $\tilde m_{1}$ can be chosen
to be zero which results in $\tilde m_2 = - m_1/\tan^2 \theta$.
Using this relation, the value of $m_2$ is determined by
\begin{eqnarray}
m^2_2 = {\Delta m^2_{sol}\over 1- \tan^4\theta}.
\end{eqnarray}

The mass $m_3$ can be expressed as
\begin{eqnarray}
&&\mbox{Normal hierarchy}:\;\; m^2_3 = \Delta m^2_{atm} +
{\Delta
m^2_{sol}\over 1- \tan^4\theta};\nonumber\\
&&\mbox{Reversed hierarchy}:\;\; m^2_3 = -\Delta m^2_{atm} +
{\Delta m^2_{sol}\over 1- \tan^4\theta}.
\end{eqnarray}
The phase of $\tilde m_3$ is not determined from the above
consideration. The reversed hierarchy is not a physical solution
in this case because $m^2_3$ is minus numerically when constraints
from data are imposed.

From eqs (9) and (10), we obtain the central value of $m_1$ for
the normal hierarchy to be 0.0038 eV and the 3$\sigma$ lower bound
to be 0.0025 eV. The sum of the masses
 $m_{sum}$ is equal to 0.062 $\pm$ 0.010
eV which satisfies the WMAP bound and can be probed by future
cosmological data.  The effective mass $m_{ee}$ is identically
equal to zero and $\langle m_\nu\rangle$ =0.0062 $\pm$ 0.0011 eV
which are safely within the current experimental bounds and is
very difficult to be probed by near future experiments.

The case L2 is a special case of the minimal Zee mass
matrix\cite{zee} and has been shown to be ruled out\cite{he}. The
cases L2  and L3 cannot be consistent with data, it is easy to
understand from the simultaneous requirements $M_{11} = 0$, and
$M_{22}=0$ ($M_{23} = 0$). Because of this, one has $\tilde m_3 =
s^2 \tilde m_1 + c^2 \tilde m_2$ and $ c^2 \tilde m_1 + s^2 \tilde
m_2 =0$ which leads to
\begin{eqnarray}
{\Delta m^2_{atm}\over \Delta m^2_{sol}} = {\tan^2\theta
(2-\tan^2\theta)\over 1- \tan^4 \theta} .
\end{eqnarray}
With $0.28 <\tan^2\theta < 0.6$, the above ratio is in the range
0.5 to 1.3 which is in conflict with data. This class of models is
therefore ruled out.

We now discuss the case L4. The constraint from $M_{22} = m_1 s^2
e^{-i2\rho_1} + m_2 c^2 e^{-i2\rho_2} + m_3 e^{-i2\rho_3} = 0$
implies that $m_1 s^2 + m_2 c^2 \geq m_3$. This implies that only
reversed mass hierarchy, $m_2 > m_1 > m_3$, is allowed. We choose
the convention where the phase $\rho_2 = 0$. We find that the
Majorana phases $\rho_{1,3}$ are determined by the masses and
mixing angles as
\begin{eqnarray}
&&\cos2\rho_1 = {1\over 2 m_1 m_2 s^2 c^2} (m^2_3 - m^2_1 s^4
-m^2_2 c^4),\nonumber\\
&&\cos2\rho_3 = {1\over m_3} (m_1s^2 \cos2\rho_1 + m_2
c^2),\nonumber\\
&&\sin2\rho_3 = {1\over m_3} m_1 s^2 \sin 2\rho_1 .\label{LL4}
\end{eqnarray}
There are two solutions, due to the undetermined sign of $\sin
2\rho_1 = \pm \sqrt{1-\cos^22\rho_1}$, even with $\cos2\rho_1$
fixed. They cannot be distinguished by oscillation and laboratory
mass measurement experiments. If leptogenesis is responsible for
the baryon asymmetry of our universe, we will show later that the
two different solutions give different signs for the baryon
asymmetry and the solution with positive $\sin2\rho_1$ has to be
chosen.

\begin{figure}[htb]
\begin{center}
\includegraphics[width=7cm]{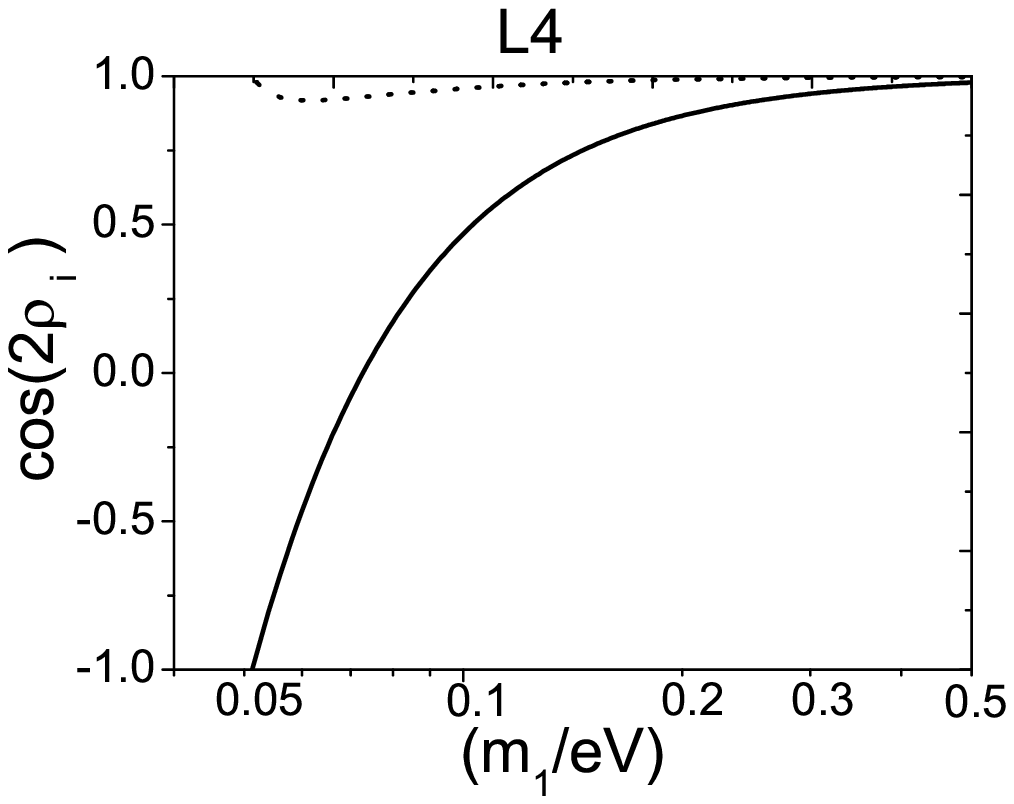}\includegraphics[width=7cm]{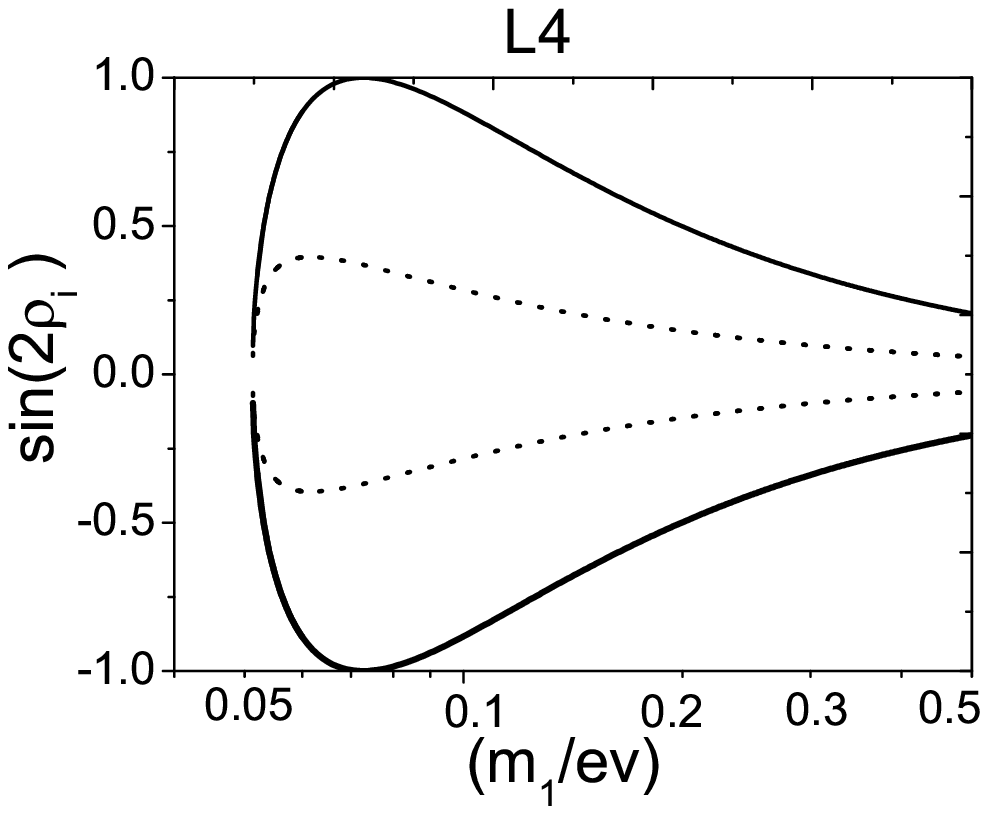}
\\(a)    \hspace{6cm}        (b)\\
\end{center}

\caption{$\cos 2\rho_1$ and $\cos 2\rho_3$ (Fig.\ref{L4} (a)), and
$\sin2\rho_1$ and $\sin2\rho_3$ (Fig.\ref{L4} (b)), respectively,
as functions of $Log(m_{1})$ for the central values of $\Delta
m^2_{solar}$, $\Delta m^2_{atm}$, and $tan \theta$. Both solutions
with $\pm |\sin 2\rho_1|$ are drawn. In (a) the solid line is for
$\cos 2\rho_1$ and the dotted line is for $\cos 2\rho_3$, while in
(b) the solid line is for $\sin 2\rho_1 $ and the dotted line is
for $\sin 2\rho_3 $ respectively.}\label{L4}
\end{figure}

In Fig. \ref{L4} (a) and (b), we show $\cos 2\rho_1$ and
$\cos2\rho_3$, and $\sin2\rho_1$ and $\sin2\rho_3$, respectively,
as functions of $Log(m_1)$ with the best fit values for $\Delta
m^2_{sol}$, $\Delta m^2_{atm}$ and $\tan\theta$. We see that in
order to have physical solutions  there is a minimal value for
$m_1$ which is about 0.05 eV for the input parameters with central
values. When errors in the input parameters are included, the
lower bound can be reduced to 0.039 eV at $3\sigma$ level. One can
also express the masses as functions of the phase $\rho_1$.

There is no upper bound for the neutrino masses from the above
considerations. If one takes the WMAP constraint, $m_1$ is bounded
to be less than 0.238 eV. This implies that $\cos2\rho_1$ is to be
smaller than 0.906.

The effective masses $m_{ee} = |M_{11}| = (m_1^2 c^4 + m^2_2 s^4 +
2m_1 m_2s^2 c^2 \cos2\rho_1)^{1/2}$ and $\langle m_\nu \rangle =
(c^2m^2_1 + s^2 m^2_2)^{1/2}$ are constrained. Compared with
current experimental upper bounds for  $m_{ee} < $ 1.35 eV and
$\langle m_{\nu}\rangle < $  3 eV , we get the upper bound for
$m_{1}$ to be 0.95 eV which is above the WMAP bound. The lower
bound for $m_{ee}$ and $\langle m_\nu \rangle$ may be calculated
too. They are, at $3\sigma$ level, 0.0217 eV and 0.0392 eV
respectively. The lower bound on $m_{ee}$ can be probed by future
neutrinoless double beta decays.

The constraints on L5 can be obtained by replacing $m_3$ by $-m_3$
in eq. (\ref{LL4}). The net result is to change the signs of $\cos
2\rho_3$ and $\sin 2 \rho_3$. Leptogenesis will select the
solution with $\sin 2\rho_1$ to be positive again.

\subsection{Constraints From Texture Zeros In The Heavy Neutrino Mass
Matrix}

We now discuss the situation for the cases in class b). The cases
R1, R2 and R3 all require $M^h_{11}=0$, which implies
\begin{eqnarray}
s^2 \tilde m_1 + c^2 \tilde m_2=0.
\end{eqnarray}

Since data requires that $s^2 <c^2$ and $m^2_2 > m_1^2$, it is not
possible to satisfy the above equation. Cases R1, R2 and R3 are
therefore ruled out by data.

A specific realization of R4 was discussed by Grimus and Lavoura
in Ref.\cite{grimus1}. We choose the convention with $\rho_2 = 0$.
In this case since $M_{23} = s^2/\tilde m_1 + c^2/\tilde m_2 -
1/\tilde m_3=0$, when combined with $m_2> m_1$ from data, one
obtains $m_3
> m_1$. Since data show that $|\Delta m^2_{32}|$ is larger than
$|\Delta m_{21}^2|$, only normal hierarchy neutrino mass pattern
is allowed.

The condition $|M_{23}|=0$ also leads to
\begin{eqnarray}
&&\cos2\rho_1 ={1 \over 2 m_{1} m_{2} c^2 s^2}( {m^2_{2}m^2_{1}
\over m^2_{3}} - m^2_{2}
s^4-m^2_{1} c^4),\nonumber\\
&&\cos 2\rho_3 =    {m_{3}\over m_{1}m_{2}} ( {m_{2}s^2 \cos(
2\rho_1)
+m_{1} c^2}),\nonumber\\
&&\sin 2\rho_3 = {m_{3}\over m_{1}}s^2\sin(2\rho_1). \label{sin}
\end{eqnarray}
Similar to the case for L4, there are two solutions due to the
undetermined sign of $\sin 2\rho_1$. Neutrino oscillation and
laboratory neutrino mass measurement experiments will not be able
to decide which solution to take. However, leptogenesis will
select the solution with positive $\sin 2 \rho_3$ .

\begin{figure}[htb]
\begin{center}
\includegraphics[width=7cm]{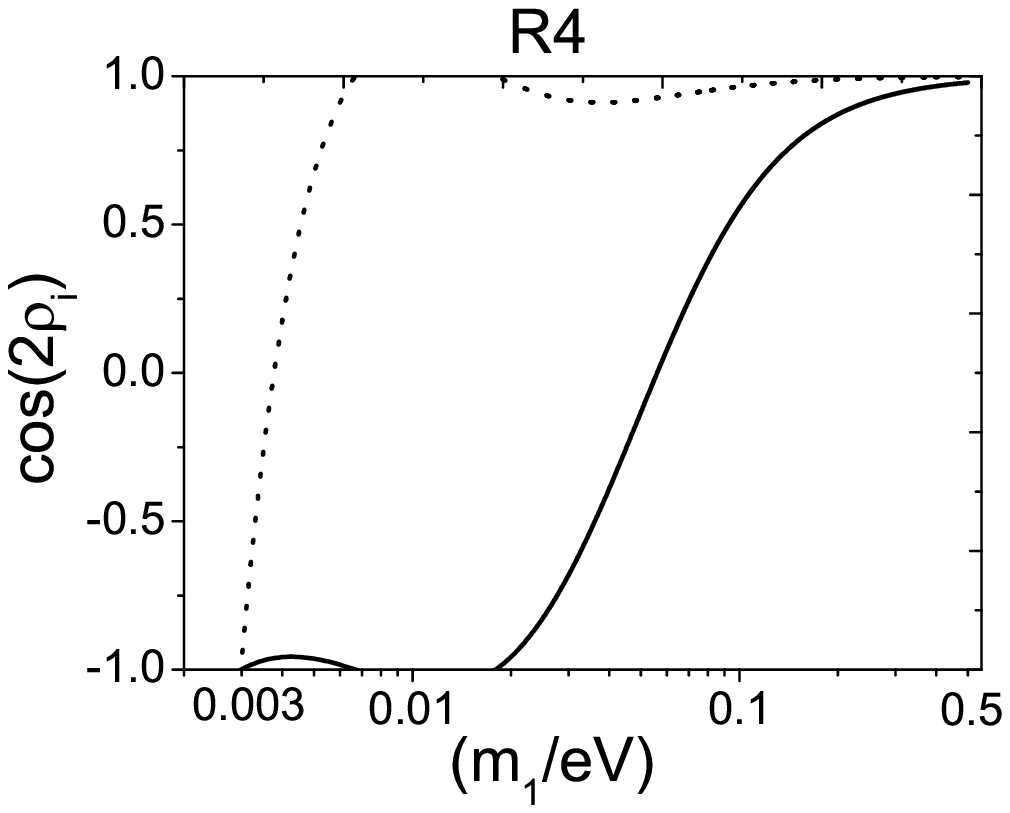}\includegraphics[width=7cm]{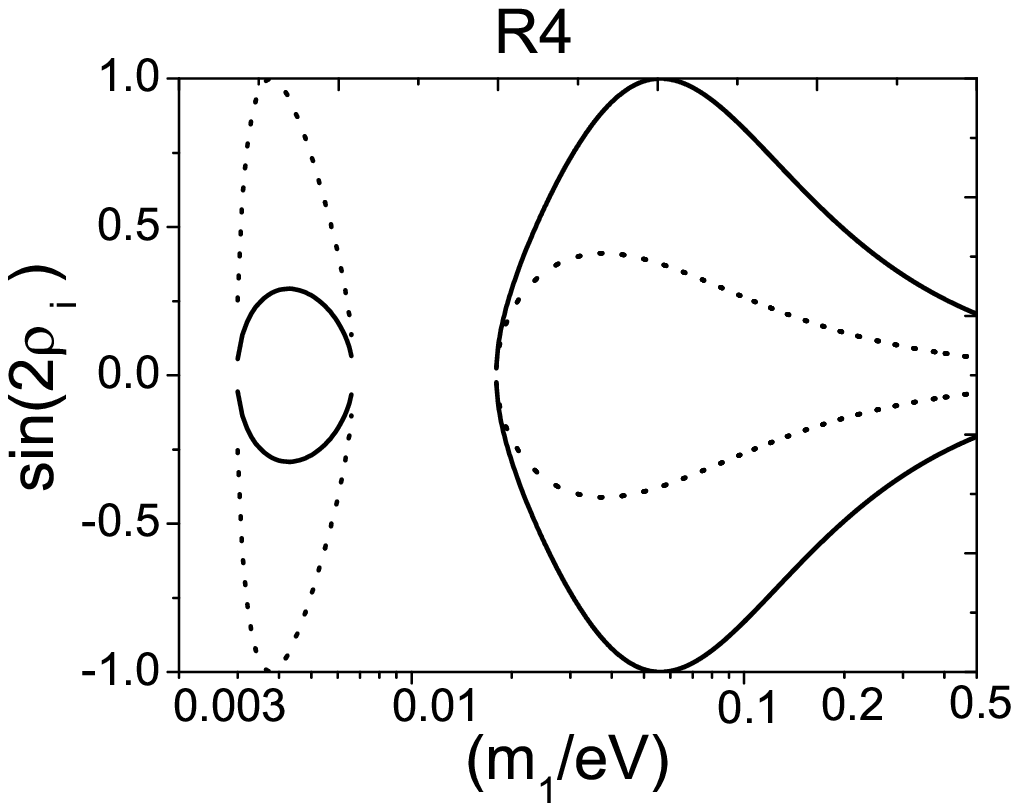}
\\(a)    \hspace{6cm}        (b)\\
\end{center}

\caption{$\cos 2\rho_1$ and $\cos 2\rho_3$ (Fig.\ref{R4} (a)), and
$\sin2\rho_1$ and $\sin2\rho_3$ (Fig.\ref{R4} (b)), respectively,
as functions of $Log(m_{1})$ for the central values of $\Delta
m^2_{solar}$, $\Delta m^2_{atm}$, and $tan \theta$. Both solutions
with $\pm |\sin 2\rho_1|$ are drawn. In (a) the solid line is for
$\cos 2\rho_1$ and the dotted line is for $\cos 2\rho_3$ ,while in
(b) the solid line is for $\sin 2\rho_1 $ and the dotted line is
for $\sin 2\rho_3 $ respectively.}\label{R4}
\end{figure}

In Fig. \ref{R4} (a), we show $\cos 2\rho_1$ and $\cos2\rho_3$ as
functions of $m_1$ with the best fit values for $\Delta
m^2_{sol}$, $\Delta m^2_{atm}$, $\tan\theta$. In Fig. \ref{R4} (b)
we show the two solutions for $\sin 2\rho_1$ and $\sin 2 \rho_3$
as functions of $Log(m_1)$. There is a region, around $m_1 =
0.01$,
 not allowed. We see that there is a minimal value for
$m_1$ which is about 0.003 eV. With errors in the input
parameters, the low bound at $3\sigma$ level is 0.002 eV. One can
also express the masses as functions of the phase $\rho_1$.

Similar to the case L4, there is no upper bound for the neutrino
masses for R4 from the above considerations. If one takes the WMAP
constraint, $\cos2\rho_1$ is bounded to be smaller than 0.905.

The effective masses $m_{ee} = |M_{11}| = m_1 m_2/m_3$ and
$\langle m_\nu \rangle = (c^2m^2_1 + s^2 m^2_2)^{1/2}$ are
constrained. Compared with current experiment upper bounds for
$m_{ee} < $ 1.35 eV and $\langle m_\nu \rangle <$ 3 eV , we get
the upper bound for $m_{1}$ to be 1.35 eV which is again above the
WMAP bound. The lower bound at 3$\sigma$ for $m_{ee}$ and $\langle
m_\nu \rangle$ are $3.8\times 10^{-4}$ eV and $4.7\times10^{-3}$
eV respectively.

Again the constraints for case R5 can be obtained by simply
changing $\tilde m_3$ to $-\tilde m_3$.

\section{Heavy Neutrino Masses and Mixing Matrix}

In our previous discussions we have concentrated only on the light
neutrino masses, mixing and phases. We have seen that the mass
matrix is completely specified by experimental measurable
quantities. In fact once the light neutrino mass matrix is known,
the right-handed neutrino mass matrix is almost specified as can
be seen from eq.(\ref{mr}).

There are three new parameters $v_\nu$, $b$ and $\epsilon$ in
$M_R$. In the cases considered here, only the combination $v_\nu
b$ appears in the calculations. We will normalize $v_\nu$ to have
the SM values of 174 GeV and let $b$ be a free parameter. It is
interesting to note that if one knows $\epsilon$, all information
on the mixing matrix $U_R$ is known, and also the ratios of the
heavy neutrino masses $M_i/M_j$ are known once the light neutrino
masses and mixing angles are fixed.

In the limit $\epsilon = 1$, $U_R =U^*$ and $M_i = v^2_\nu
b^2/m_i$. When $\epsilon$ is not equal to 1, the situation is more
complicated. But from eq. (\ref{mr}) it is clear that the unitary
matrix $U_R$ which diagonalizes $M_R$ still has the bi-large
mixing form. When $\epsilon$ is close to one, the heavy neutrino
mass hierarchies are $M_3 > M_1 > M_2$ and $M_1 > M_2
> M_3$ for the reversed and normal light neutrino hierarchies,
respectively. When $\epsilon$ deviates from one, the mass
hierarchy pattern will change and $U_R$ is non-trivially related
to $U$. But $M_3$ is always equal to $v_\nu^2b^2/\tilde m_3$.

\begin{figure}[htb]
\begin{center}
\includegraphics[width=5cm]{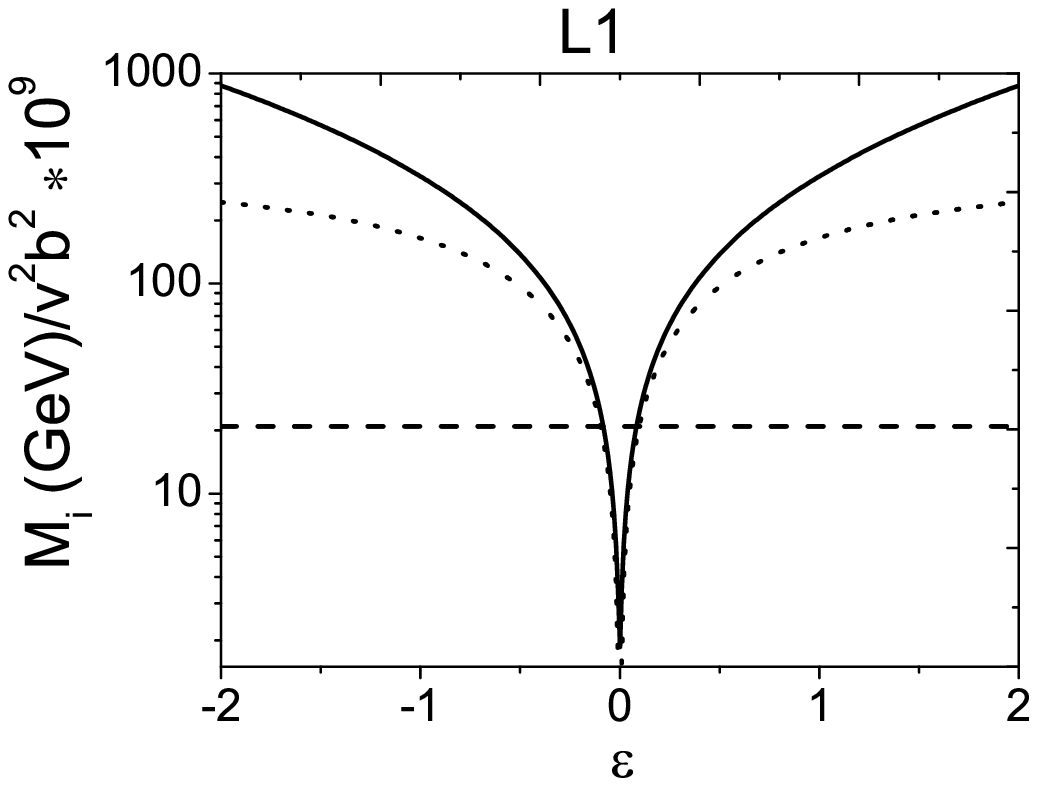}
\\
(a)
\\\vspace{0.5cm}
\includegraphics[width=5cm]{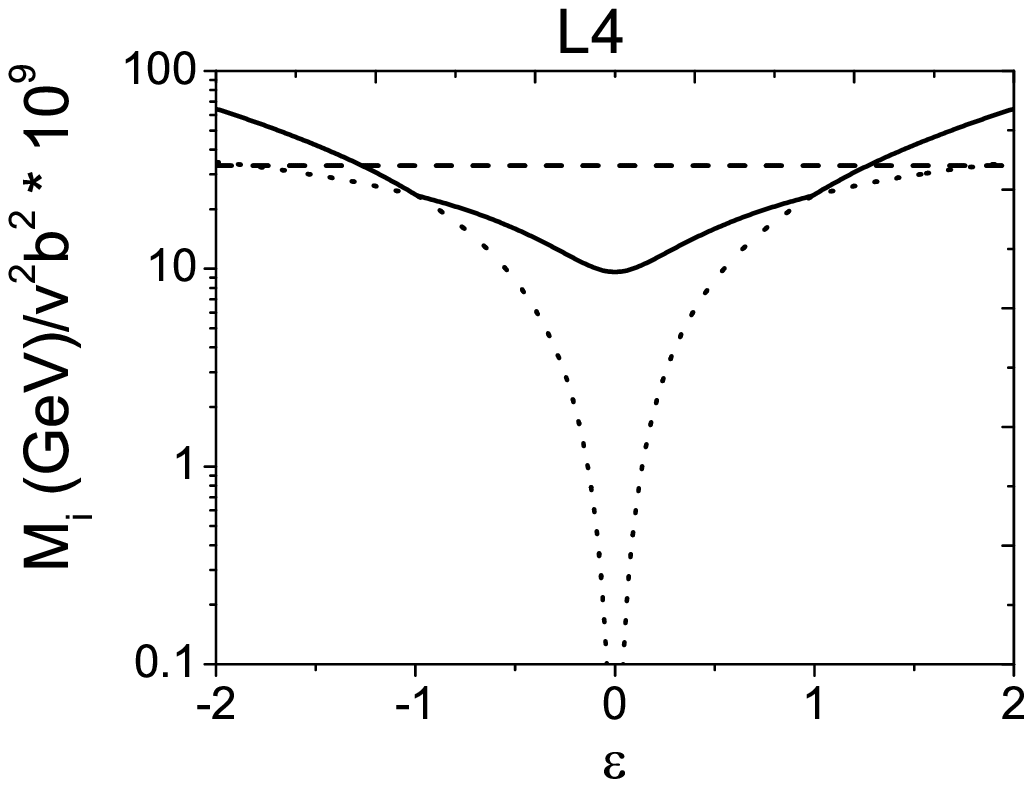}
\includegraphics[width=5cm]{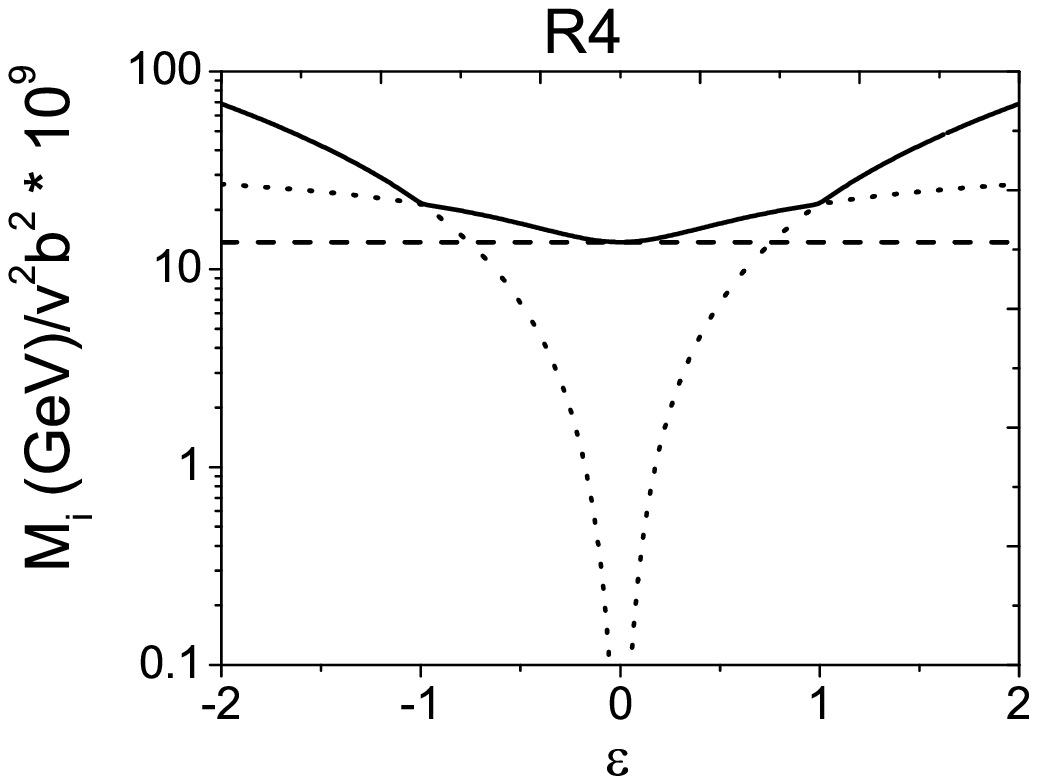}
\\
(b)
\\\vspace{0.5cm}
\includegraphics[width=5cm]{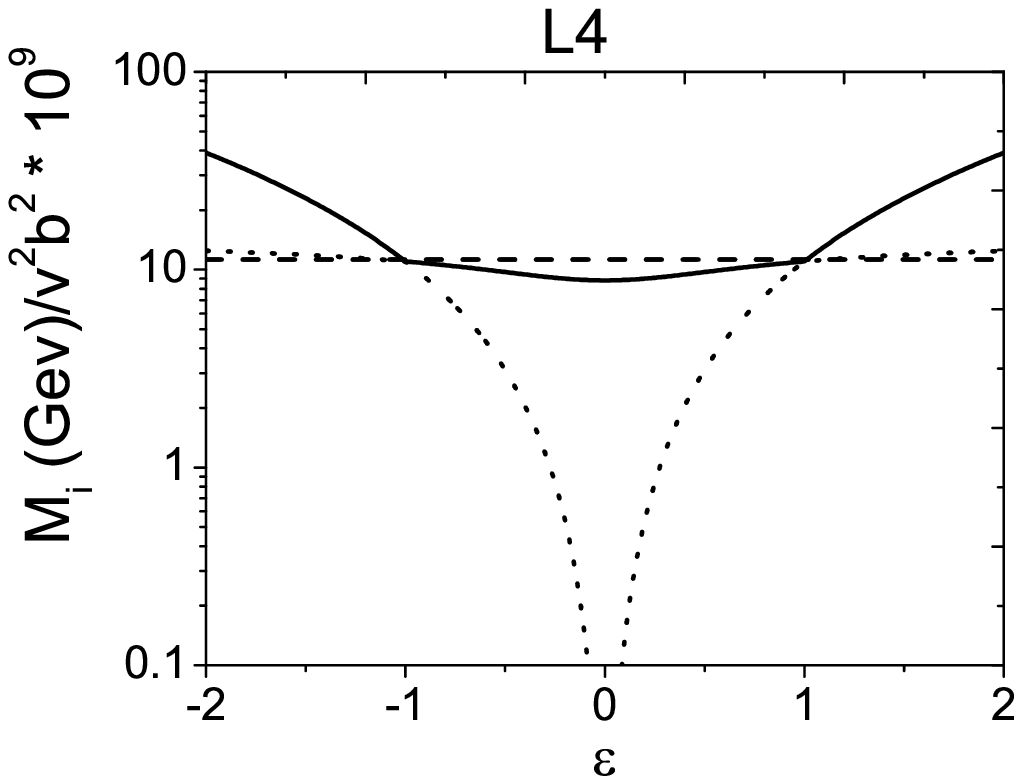}
\includegraphics[width=5cm]{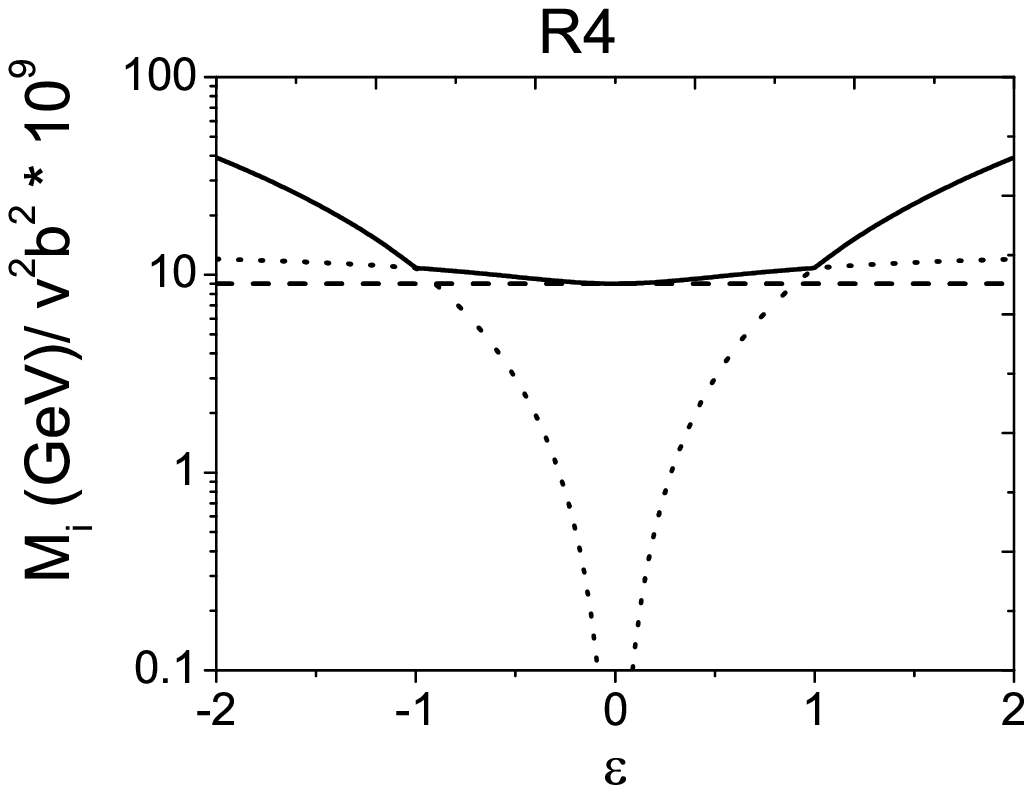}
\\
(c)
\end{center} \caption{$M_i$(GeV)/$v^2_\nu b^2$ for cases L1, L4 and
R4 as functions of $\epsilon$. For L1, $m_1$ is equal to 0.0038 eV
((a)) (determined from the central values of the mixing angles).
For L4 and R4, $m_1$ is not fixed by the mixing angles. We draw
figures for $m_1 = 0.055$ eV ((b)) and $m_1 = 0.1$ eV ((c)) for
illustrations. The solid, dotted and dashed lines are for $M_1$,
$M_2$ and $M_3$, respectively.}\label{Mi}
\end{figure}

In Fig.\ref{Mi}, we show the heavy neutrino masses as functions of
$\epsilon$ for  several fixed values of $m_1$ for illustration.
For case L1, we use the central value 0.0038 eV, and for L4 and R4
we use two typical values 0.055 eV and 0.1 eV for $m_1$,
respectively. The cases L4 and L5 have the same eigen-masses, and
R4 and R5 also have the same eigen-masses. From Fig.\ref{Mi} we
can clearly see that the mass hierarchy changes with $\epsilon$.

The mixing matrix $U_R$ is more complicated. It has
the general form
\begin{eqnarray}
U_R = \left ( \begin{array}{lll}
e^{i\delta_1}&0&0\\
0&e^{i\delta_2}&0\\
0&0&e^{i\delta_2}
\end{array}
\right )
\left (\begin{array}{rrr}
c'&s'&0\\
-{s'\over \sqrt{2}}&{c'\over \sqrt{2}}&{1\over \sqrt{2}}\\
-{s'\over \sqrt{2}}&{c'\over \sqrt{2}}&-{1\over \sqrt{2}}
\end{array}\right )\left ( \begin{array}{lll}
1&0&0\\
0&e^{i\gamma_2}&0\\
0&0&e^{i\gamma_3}
\end{array}
\right ).
\end{eqnarray}

In the following we give, as examples, $U_R$ for  the cases L1, L4
and R4. In the case L1 with $m_1 = 0.0038$ eV,

\begin{eqnarray}
U_R = \left (\begin{array}{ccc}
0.7175& 0.6965 & 0\\
-0.4925 &  0.5074 & 0.707\\
-0.4925 & 0.5074 & -0.707
\end{array}
\right ).
\end{eqnarray}

For cases L4 and R4, in the basis where all eigen-masses are real
and positive, we write $U_R$ for two typical values of $m_1$, $
0.055$ eV and $0.1$ eV. We have

For case L4 with $m_1 = 0.055$ eV,
\begin{eqnarray}
U_R=\left ( \begin{array}{ccc}
0.2946+0.1841\ i&-0.4986+0.7942\ i&0\\
0.6516-0.1225\ i&-0.0449-0.2415\ i&-0.6941+0.1351\ i\\
0.6516-0.1225\ i&-0.0449-0.2415\ i&0.6941-0.1351\ i
\end{array}
\right ),
\end{eqnarray}
and with $m_1 = 0.1$ eV,

\begin{eqnarray}
U_R=\left ( \begin{array}{ccc}
0.0532+0.1285\ i& -0.9162+0.3760\ i& 0\\
0.6929-0.1012\ i& -0.0139-0.0973\ i& -0.6996+0.1029\ i\\
0.6929-0.1012\ i& -0.0139-0.0973\ i& 0.6996-0.1029\ i
\end{array}
\right ).
\end{eqnarray}

For case R4 with $m_1= 0.055$ eV,

\begin{eqnarray}
U_R=\left ( \begin{array}{ccc}
0.1202+0.1778\ i&-0.8122+0.5424\ i&0\\
0.6777-0.1332\ i&-0.0284-0.1491\ i&-0.6938+0.1364\ i\\
0.6777-0.1332\ i&-0.0284-0.1491\ i&0.6938-0.1364\ i
\end{array}
\right ),
\end{eqnarray}
and with $m_1=0.1$ eV,

\begin{eqnarray}
U_R=\left ( \begin{array}{ccc}
0.0431+0.1172\ i&-0.9324+0.3393\ i&0\\
0.6954-0.0931\ i&-0.0114-0.0875\ i&-0.7009+0.0938\ i\\
0.6954-0.0931\ i&-0.0114-0.0875\ i&0.7009-0.0938\ i
\end{array}
\right ).
\end{eqnarray}

\section{Leptogenesis}

There are extensive discussions on implications of leptogenesis
for See-Saw neutrino mass
matrix\cite{fukugita,leptogenesis,asy,grimus2,leptocp}. With a
general See-Saw mass matrix, it has been shown that there is
enough room in parameters space to reproduce the observed
BAU\cite{fukugita,leptogenesis,asy}. There are also more
restrictive forms of mass matrix with texture zeros which can also
reproduce the observed BAU\cite{leptocp,grimus2}. The mass
matrices discussed in the previous sections are a class of very
restrictive matrices, in particular that there is no CKM like CP
violating phase. It is interesting to see if such models can also
produce the observed BAU. We find that although there is no CKM
like CP violating phase, the required CP violation can come from
the Majorqana phase. There is a large parameter space with which
BAU can be reproduced. Taking leptogenesis as a requirement, we
show that interesting constraints on the scale of the right-handed
neutrino can be obtained. We now proceed to provide more details.

The baryon number asymmetry problem, why our universe is dominated
by matter, is one of the most outstanding problems in modern
physics. This problem is related to the ratio $\eta_B =
n_B/n_\gamma$. Here $n_B$ is the baryon number density and
$n_\gamma$ is the photon number density. If the universe contains
equal matter and anti-matter initially with baryon number
conserved, the expected ratio for $\eta_B$ is about $10^{-20}$.
Observations from Big-Bang Nucleosynthesis (BBN) and Cosmic
Macrowave Background (CMB) radiation determine $\eta_B$ to
be\cite{pdg,bbn,wmap} $6.5^{+0.4}_{-0.3}\times 10^{-10}$. There is
a huge difference between the expected and the observed values.
Sakharov showed that if there are\cite{skharov}: 1) baryon number
violation, 2) C and CP violation, and 3) occurrence of non-thermal
equilibrium
 when 1) and 2) are effective, it is possible to create
a matter dominated universe from a symmetric one in the early
epoch of the universe.

In the Standard Model due to $SU(2)_L$ anomaly, there are baryon
number violating interactions. This interaction becomes strong at
high temperatures\cite{sphelaron}. This interaction violates
$B+L$, but conserves $B-L$. Fukugita and Yanagida\cite{fukugita}
noticed that if in the early universe there was lepton number
asymmetry, this interaction can transfer lepton number asymmetry
$a_i$ produced by heavy neutrino decays, for example, to baryon
number asymmetry.

The surviving baryon asymmetry from lepton number asymmetry due to
the ``lth'' heavy neutrino is given by\cite{fukugita,th}
\begin{eqnarray}
\eta_B = \left .{s\over n_\gamma}\right |_0 {\omega\over \omega
-1} {a_l \kappa_l\over g_{*l}},\label{etab}
\end{eqnarray}
where $s=(2\pi^2/45) g_{*0} T^3|_0$ and $n_\gamma = (2/\pi^2)
\zeta (3) T^3|_0$ are the entropy and photon densities of the
present universe with $g_{*0} = 43/11$ being the effective
relativistic degrees of freedom. The parameter $\omega$ is
calculated to be\cite{th} $\omega =(8N_F + 4N_H)/(22N_F + 13 N_H)$
depending on the number of $SU(2)_L$ doublet Higgs scalars $N_H$
and fermions $N_F$. $g_{*l}$ is the effective relativistic degrees
of freedom at the temperature where the lepton number asymmetry
$a_l$ is generated from the ``lth'' heavy neutrino decay. For the
lightest heavy neutrino decay contribution, $g_{*l} =
(28+(7/8)\times 90)_{SM} + 4(N_H-1) +2(7/8)$ is of order 100. Here
the last term comes from the lightest heavy Majorana neutrino
which produces the lepton number asymmetry. The number $N_H$
depends on the details of the specific model. We have checked the
sensitivity of $\eta_B$ on $N_H$ and find that there is only about
a 10\% reduction for $N_H$ varying from 1 to 5. We will assmue
that there is just one Higgs doublet in our numerical
calculations. $\kappa_l$ is a dilute factor which depends on the
ratio of heavy Majorana neutrino decay rate and the Hubble
parameter at the time of heavy neutrino decay, $K_l =
\Gamma_l/H_l$ with $\Gamma_l = (\hat Y_\nu \hat
Y^\dagger_\nu)_{ll} M_l/8\pi$ and $H_l =
1.166\sqrt{g_{*l}}M^2_l/M_{planck}$. Here $\hat Y_\nu = V^T_R
Y_\nu$ is the Yukwawa coupling in the basis where $M_R$ is
diagonalized.

The heavy neutrino mass is of order $M_l  \sim (v^2_\nu/ m_{3})
(\hat Y_\nu \hat Y^\dagger_\nu)_{ll}$, one would obtain
$\Gamma_l/H_l \sim 10^4 (m_{3}/eV) (100 GeV/v_\nu)^2$. For $m_3$
within the allowed lower bound discussed earlier and upper bound
from WMAP, the factor $K_l$ is within the range of 10 $\sim 10^6$
. In this range the dilute factor $\kappa_l$ is approximated
by\cite{asy} $\kappa_l \approx 0.3 /K_l (\ln K_l)^{3/5}$. In our
numerical calculations we will use this approximate form.

We now study $a_i$ in the models considered. The lepton number
asymmetry $a_i$ generated by the ``ith'' heavy neutrino is given
by\cite{fukugita,leptogenesis}
\begin{eqnarray}
a_i \approx -{1\over 8 \pi} {1\over [\hat Y_\nu \hat Y^\dagger_\nu]_{ii}}
\sum_j Im\{[\hat Y_\nu \hat Y^\dagger_\nu]^2_{ij}\} f\left ({M^2_j\over M^2_i}\right ),
\end{eqnarray}
where
\begin{eqnarray}
f(x) = \sqrt{x} ({2\over x-1} + \ln{1+x\over x}).
\end{eqnarray}

Applying the above equation to the models discussed in the
previous section, we obtain the lepton number asymmetries due to
heavy neutrino decays to be
\begin{eqnarray}
a_i&=& - {1\over 8 \pi} b^2(\epsilon^2-1)^2 \sum_j Im(U_{R1i}U_{R1ij}^*)^2
{f(M^2_j/M_i^2)\over 1+ (\epsilon^2-1)|U_{R1i}|^2)},\nonumber\\
a_1 &=& - {1\over 8 \pi} b^2(\epsilon^2-1)^2 Im(U_{R11}U_{R12}^*)^2
{f(M^2_2/M_1^2)\over 1+ (\epsilon^2-1)|U_{R11}|^2)},\nonumber\\
a_2 &=& - {1\over 8 \pi} b^2(\epsilon^2-1)^2 Im(U_{R12}U_{R11}^*)^2
{f(M^2_1/M_2^2)\over 1+ (\epsilon^2-1)|U_{R12}|^2)},\nonumber\\
a_3&=&0,
\end{eqnarray}
with $Im(U_{R11}U^*_{R12})^2 = - c'^2s'^2 \sin(2\gamma_2)$. In the
above we have used the fact that $U_{R13}=0$. Note that for
$\epsilon =1$, no lepton number asymmetry can be generated.

From eq.(\ref{etab}) we see that only solutions which generate
negative $a_i$ can be candidate producing the right sign for
baryon asymmetry. This criterion selects out solutions obtained in
Section II which are not able to be distinguished by low energy
experimental data.

Several studies of leptogensis with bi-large nuetrino mixing
matrix have been carried out\cite{grimus2}. Here we follow similar
strategy to systematically study the models discussed earlier. To
demonstrate that the See-Saw model discussed here can indeed
explain the observed baryon number asymmetry, in the following we
consider a simple case with large hierarchical structure for the
heavy Majorana neutrino mass.  In this case the dominant
contribution to the surviving baryon asymmetry is from the
lightest heavy neutrino decay. The heavy neutrino with mass of
$M_3$ does not produce a non-zero asymmetry, it cannot be the
lightest heavy neutrino since it will washout baryon asymmetries
produced by the other two heavier ones in our case. One needs to
work in the parameter space where $M_3$ is not the smallest. Large
$\epsilon$ tends to make $M_{1,2}$ bigger, while does not affect
$M_3$ as can be seen from Fig. \ref{Mi}. Therefore leptogenesis
favors small $\epsilon$. Numerically we find that with
$|\epsilon|$ less than around 0.5 the lightest heavy neutrino mass
is $M_2$ and the mass squared is at least 10 times smaller than
$M^2_3$ as can be seen from Fig.\ref{Mi}. In this range, the
washout effect of the CP conserving decay of the heavy neutrino of
mass $M_3$ would be small. We will present our results for the
baryon asymmetry produced by the lightest heavy neutrino with
$\epsilon$ satisfying the condition that the lightest heavy
neutrino mass squared is at least 10 times smaller than the next
lightest heavy neutrino mass squared

For the case L1, $U_{R11}U^*_{R12}$ is real. This leads to zero
lepton asymmetry $a_l$. This type of models cannot explain the
baryon number asymmetry in the universe.

\begin{figure}[htb]
\begin{center}
\includegraphics[width=8cm]{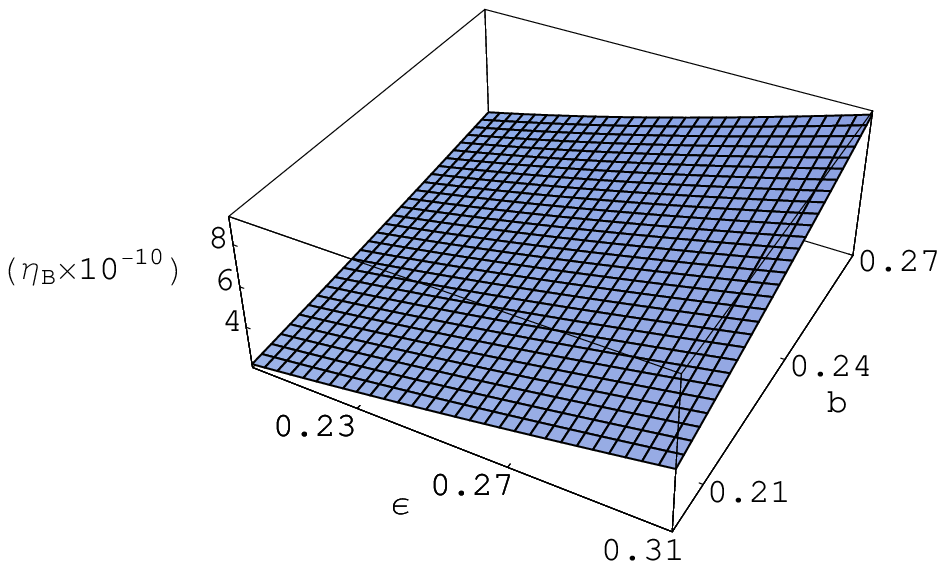}\includegraphics[width=8cm]{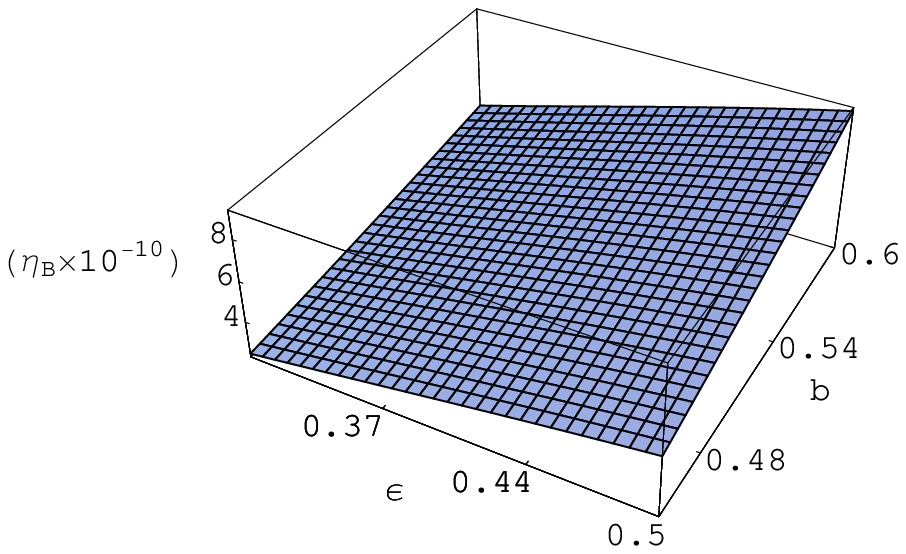}
\\(a) \hspace{6cm} (b)\\
\end{center}
\caption{ The allowed ranges for $\epsilon$ and $b$ for case L4
with $\eta_{B}$ in the range of $4\times 10^{-10} \sim 8\times
10^{-10}$ , with (a) $m_1 =0.055 $ eV and (b) $ m_1 =0.1 $ eV.
}\label{etaL4}
\end{figure}

\begin{figure}[htb]
\begin{center}
\includegraphics[width=8cm]{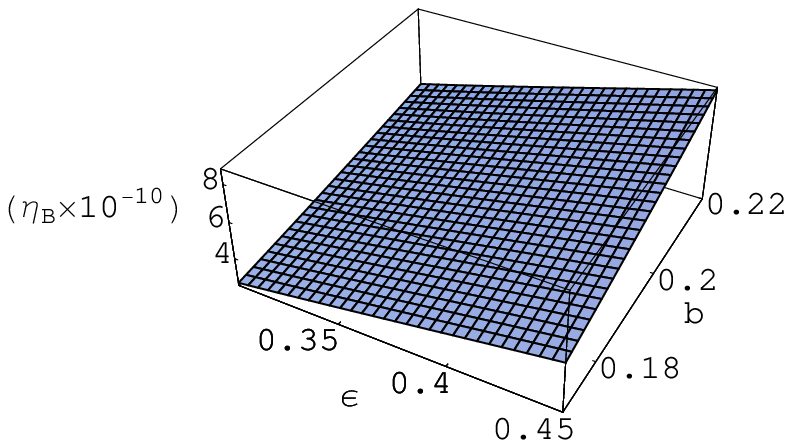}\includegraphics[width=8cm]{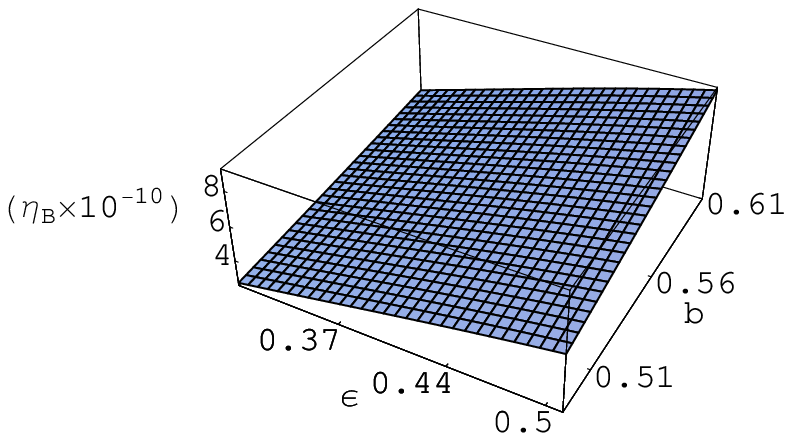}
\\(a)  \hspace{6cm}  (b)\\
\end{center}
\caption{ The allowed ranges for $\epsilon$ and $b$ for case R4
with $\eta_{B}$ in the range of $4\times 10^{-10} \sim 8\times
10^{-10}$, with (a) $m_1 =0.055 $ eV and (b)
 $m_1=0.1 $ eV .}\label{etaR4}
\end{figure}

\begin{figure}[htb]
\begin{center}
\includegraphics[width=8cm]{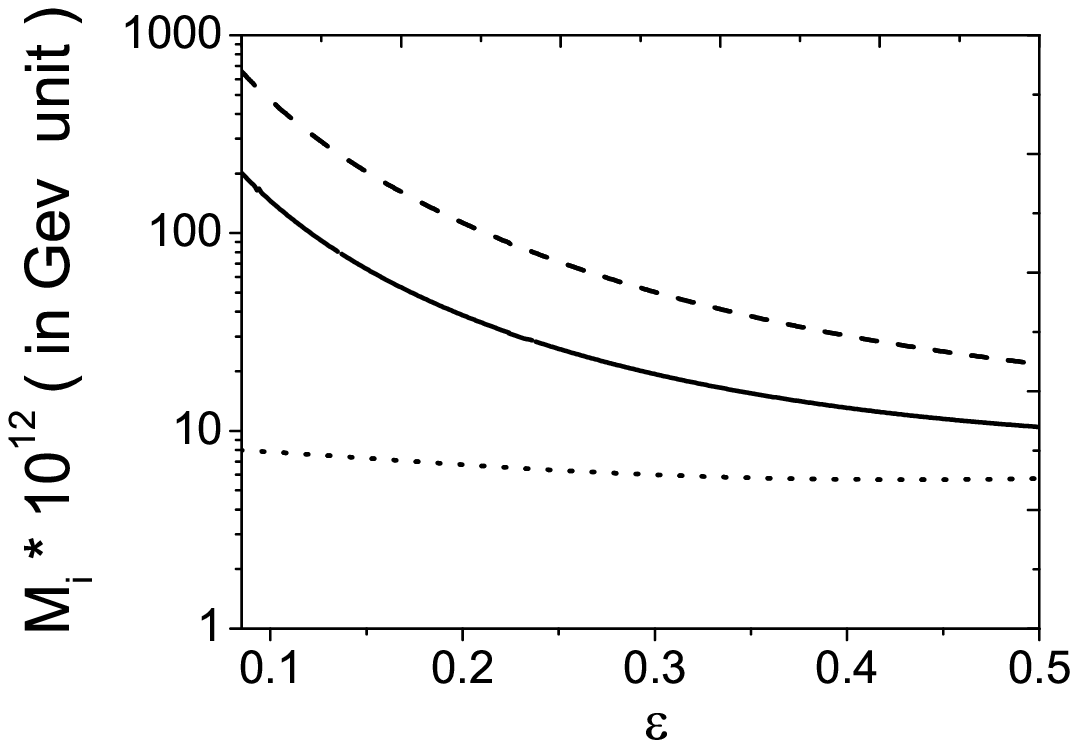}\includegraphics[width=8cm]{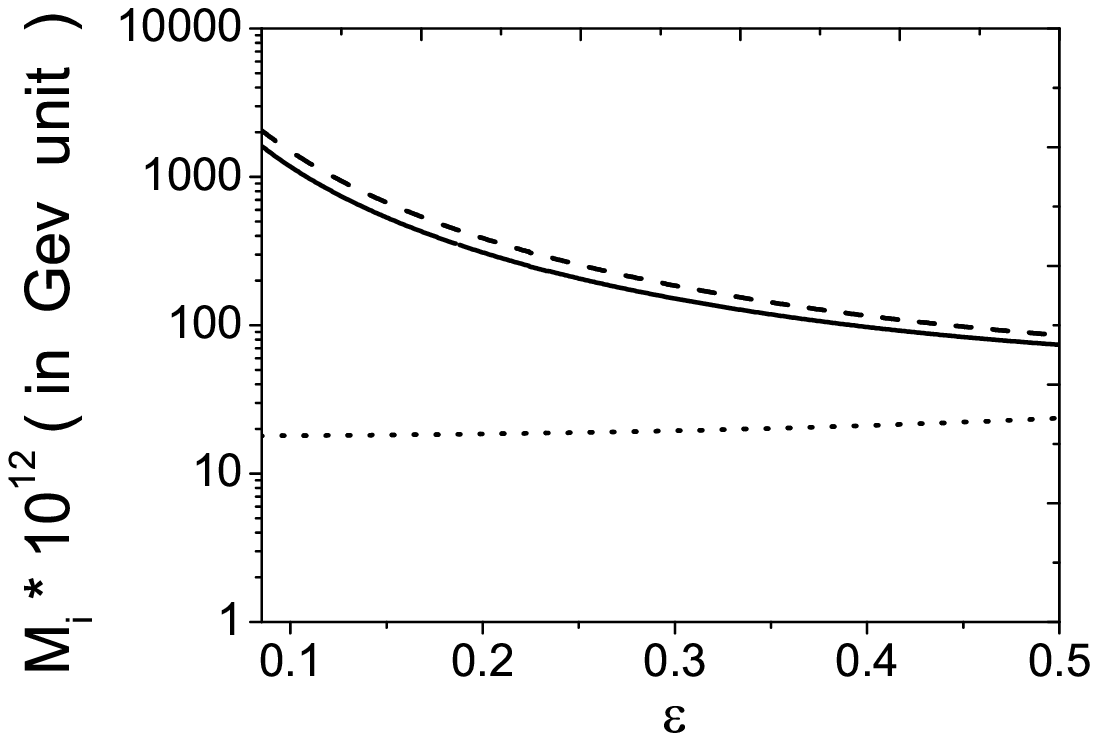}
\\(a) \hspace{6cm}  (b)\\
\end{center}
\caption{ $M_i$ for case L4  as  functions of  $\epsilon$  with
(a) $m_1 = 0.055$ eV  and (b) $m_1=0.1$ eV . The solid, doted and
dashed lines are for $M_1$, $M_2$ and $M_3$
respectively.}\label{etamiL4}
\end{figure}

\begin{figure}[htb]
\begin{center}
\includegraphics[width=8cm]{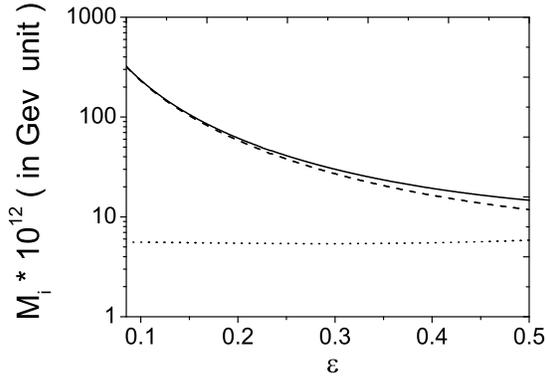}\includegraphics[width=8cm]{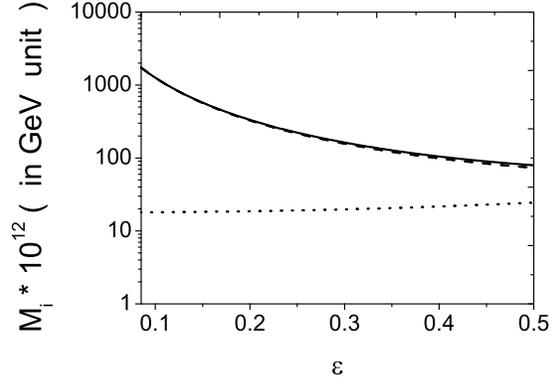}
\\(a)  \hspace{6cm}  (b)\\
\end{center}
\caption{ $M_i$ for case R4  as functions of $\epsilon$ with (a)
$m_1 = 0.055$ eV and (b) $m_1 = 0.1$ eV. The solid, doted and
dashed lines are for $M_1$, $M_2$ and $M_3$
respectively.}\label{etamiR4}
\end{figure}

In Figs.\ref{etaL4} and \ref{etaR4}, we show $\eta_B$ as functions
of b and $\epsilon$ for $m_1 = 0.055$ eV and $0.1$ eV for the
cases L4 and R4. Only the cases with negative $a_i$ which produces
the right sign for the observed baryon number asymmetry are shown.
There are two solutions with different signs for $\sin 2\rho_1$ in
the case of L4 which satisfies neutrino mass and oscillation
experimental constraints as discussed in section II. If the model
is required to produce the baryon number asymmetry, we find that
only the solution with the positive $\sin 2\rho_3$ is allowed.
Similar situation happens for the case of R4, positive $\sin 2
\rho_1$ has to be chosen. For the cases L5 and R5, the solutions
with positive $\sin 2\rho_3$ have to be chosen.

We see from Figs. \ref{etaL4} and \ref{etaR4} that the observed
baryon number asymmetry can be produced in the models considered
here. We also see that the requirement of generating the correct
baryon number asymmetry, the parameters $\epsilon$ and $b$ are
constrained. One can use this fact to obtain the allowed mass
ranges for the heavy neutrino masses $M_i$. In Figs. \ref{etamiL4}
and \ref{etamiR4} we show $M_i$ as functions of $\epsilon$ for the
central value of $\eta_B$. These masses represent possible new
physics scale and are constrained to be in the range of $10^{12}
\sim 10^{15}$ GeV.

\section{Conclusions}

In this paper we have studied constraints from texture zeros in
bi-large mixing See-Saw neutrino mass matrices and also from
leptogenesis. We have systematically investigated two classes of
models with one of them (class a)) to have the texture zeros
imposed on the light neutrino mass matrix, and another (class b))
to have the texture zeros imposed on the heavy neutrino mass
matrices.

Assuming a simple form proportional to $diag(\epsilon, 1, 1)$ for
the Dirac mass matrix which relates the left- and right- handed
neutrinos, both light and heavy neutrinos can simultaneously have
the bi-large mixing matrix form. Both classes a) and b) of mass
matrices can have 5 different forms which produce non-trivial
three generation mixing. We find that only three (L1, L4, L5 ) in
class a) and two (R4, R5) in class b), respectively, can be
consistent with present data on neutrino masses and mixing
constraints. In all the models none of the neutrino masses can be
zero. Using present data, the lightest neutrino is bounded to be
heavier than, 0.0025 eV, 0.039 eV and 0.002 eV for L1, L4 and L5,
and, R4 and R5, respectively. Future experiments can provide
further tests and even rule out some of the models.

Because $V_{13}=0$, there is no CKM type of CP violating phase in
the light neutrino mixing matrix.  No CP violating effects can be
observed in neutrino oscillation experiments. However, there can
be non-trivial Majorana phases. These phases can play an important
role in explaining the observed baryon number asymmetry in our
universe. We have shown that in the models considered there are
parameter spaces where the observed baryon number asymmetry can
indeed be generated through the leptogenesis mechanism. It is
interesting to note that the requirement of producing the observed
baryon number asymmetry rules out several models which are,
otherwise, impossible to achieve by Laboratory experiments. This
requirement also provides a condition to fix the allowed scale for
the heavy neutrinos. We find that the masses are in the range of
$10^{12} \sim 10^{15}$ GeV.

In the models we considered $V_{13}$ is zero which is allowed by
present experimental data and can be tested by future experiments.
Several experiments are planned to measure $V_{13}$ with greater
precision\cite{daya}. Obviously should a non-zero value for
$V_{13}$ be measured, modifications for the model considered are
needed. However, the models considered can be taken as the lowest
order approximations. How to obtain such mass matrices deserves
more future theoretical studies.

\acknowledgments This work was supported in part by grants from
NNSFC and NSC. XGH thanks the College of Physics at Nankai
University for hospitality where part of the work was carried out.



\end{document}